\documentclass{PoS}

\usepackage[figuresright]{rotating}

\title{The effect of r-process enhancement in binary CEMP-s/r stars}

\ShortTitle{CEMP-s/r stars}

\author{\speaker{S.~Bisterzo}\\%
        Dipartimento di Fisica Generale, Universit$\grave{\rm a}$ di Torino, via P. Giuria 1, 10025  Torino, Italy\\
        E-mail: \email{bisterzo@ph.unito.it}}

\author{R.~Gallino\\
        Dipartimento di Fisica Generale, Universit$\grave{\rm a}$ di Torino, via P. Giuria 1, 10025 Torino, Italy\\
        INAF Osservatorio Astronomico di Collurania, via M. 
   Maggini, 64100 Teramo, Italy\\
        E-mail: \email{gallino@ph.unito.it}}

\abstract{About half of carbon and $s$-process enhanced metal-poor 
stars (CEMP-$s$) show a high $r$-process enrichment (CEMP-$s/r$), 
incompatible with a pure $s$-process contribution.
CEMP-$s$ stars are of low mass ($M$ $<$ 0.9 $M_\odot$) and belong to binary 
systems. The C and $s$-process enrichment 
results from mass transfer by the winds of the primary AGB companion 
(now a white dwarf). The nucleosynthesis of the $r$-process, instead, is believed
to occur in massive stars exploding as Supernovae of Type II.
The most representative $r$-process element is Eu (95\% of solar
Eu).
\\
We suggest that the $r$-process enrichment was already present by local SNII pollution in 
the molecular cloud from which the binary system formed. 
The initial $r$-enrichment does not affect the $s$-process nucleosynthesis.
However, the $s$-process indicators [hs/ls] (where ls is defined as the average of 
Y and Zr; hs as the average of La, Nd, Sm) and [Pb/hs] may depend on
the initial $r$-enhancement.
For instance, the hs-peak has to account of an $r$-process contribution 
estimated to be 30\% for solar La, 40\% for solar Nd, and 70\% for solar Sm.
A large spread of [Eu/Fe] is observed in unevolved halo stars up to [Eu/Fe] $\sim$ 2.
 In presence of a very high initial $r$-enrichment of the molecular cloud,
  the maximum [hs/Fe] predicted in CEMP-$s/r$ stars may increase
up to 0.3 dex. 
Instead, the spread of [Y,Zr/Fe] observed in unevolved halo stars reaches a maximum
of only $\sim$ 0.5 dex, not affecting much the predicted [ls/Fe].
This is in agreement 
with observations of CEMP-$s/r$ stars that show an observed [hs/ls] in average 
higher than that observed in CEMP-$s$. Preliminary results are presented.}

\FullConference{11th Symposium on Nuclei in the Cosmos\\
                 19-23 July 2010 \\
                 Heidelberg, Germany.}

\begin{document}

\section{Introduction}

It is commonly believed that the $s$- and $r$-processes derive
from separate astrophysical sites \cite{burbidge57}. 
The nucleosynthesis of the $s$-process occurs in stars of low mass 
(1.3 $\leq$ $M/M_\odot$ $\leq$ 8) during their thermally pulsing asymptotic 
giant branch (TP-AGB) phase. 
The main neutron source is the $^{13}$C($\alpha$, n)$^{16}$O, which burns
radiatively at $T$ $\sim$ 0.9 $\times$ 10$^8$ K during the interpulse period
in the region between the H- and He-shell (He-intershell). 
A second neutron source, $^{22}$Ne($\alpha$, n)$^{25}$Mg, is marginally 
activated at the bottom of the recurrent convective thermal instability
(thermal pulse, TP) in the He-intershell, 
mainly affecting the abundance at the branching points that 
are sensitive to temperature and neutron density.
The $s$-process elements are mixed with the surface during
the third dredge-up (TDU) episodes, in which the convective envelope engulfs 
part of the He-intershell, after the quenching of a TP.
We refer to the reviews by \cite{busso99,kaeppeler10} 
for major details on the AGB nucleosynthesis.
\\
Instead, the physical environment for the $r$-process is still unknown, 
although SNII are the most promising candidates. 
For elements from Ba to Bi, observations of very metal-poor stars with 
high $r$-enhancement (e.g., CS 22892--052 \cite{sneden03}) 
show an abundance distribution well reproduced by a scaled solar $r$-process 
residual contribution \cite{arlandini99}.
Instead, lighter neutron capture elements with Z $\leq$ 47 show values lower
than the scaled solar-system $r$-process \cite{wasserburg96,SCG08}.
The nucleosynthesis site(s) and the exact 
contributions from different primary processes to Sr, Y, Zr is highly debated \cite{travaglio04,farouqi10,pignatari10,qian08,montes07}, although also related
to massive stars.
A large spread is observed for [Eu/Fe] and for [Sr,Y,Zr/Fe] in unevolved 
halo stars. For [Fe/H] $<$ $-$2, different ranges are observed for Eu and Sr, Y, Zr:
 $-$1 $\leq$ [Eu/Fe] $\leq$ 2 with an average around 0.5 dex, while 
$-$1 $\leq$ [Sr,Y,Zr/Fe] $\leq$ 0.5 with an average around 0 dex \cite{travaglio04,montes07}.
This may be interpreted as a signature of incomplete mixing 
in the gas cloud from which these stars have formed \cite{ishimaru99,travaglio01},
as well as an indication of different and uncorrelated 
primary process contributions.

In the last years, a quite large number of 
carbon and $s$-process 
enhanced metal-poor (CEMP-$s$) stars have been detected.
CEMP-$s$ are main-sequence/turnoff or giants of low mass ($M$ $<$ 
0.9 $M_\odot$). The most plausible explanation for their peculiar high 
$s$-element abundances is mass transfer by stellar winds from the most 
massive AGB companion (now a white dwarf).
About half of these CEMP-$s$ stars are also highly enhanced in $r$-process 
elements (CEMP-$s/r$). The observed $r$-enhancement in these stars reflects 
the observations of unevolved Galactic stars at low metallicity.
CEMP-$s/r$ stars show abundance patterns incompatible 
with a pure $s$-process nucleosynthesis. While a pure $s$-process
predicts [La/Eu] $\sim$ 0.8 -- 1.1 (where La and Eu are typical $s$- and 
$r$-process elements, respectively), CEMP-$s/r$ stars show 0.0 $\leq$ [La/Eu]
$\leq$ 0.4, with [La/Fe] and [Eu/Fe] up to $\sim$ 2 dex. 
Different scenarios have been proposed in the literature to explain the origin 
of CEMP-$s/r$ (e.g., \cite{jonsell06,cohen03}). 
\\
We suggest that the molecular cloud from which the binary system formed
was already enriched in $r$-process elements by local pollution of
SNII ejecta \cite{SCG08,bisterzo09pasa}.
This hypothesis is supported
by numerical simulations by \cite{vanhala98}, who found that SNII
explosion in a molecular cloud may trigger the formation
of binary systems.
These simulations may explain the very high fraction of CEMP-$s/r$ ($\sim$ 
50\%) among the CEMP-$s$.

We present here a preliminary analysis of a 
comparison between AGB theoretical predictions and spectroscopic observations 
of CEMP-$s$ and CEMP-$s/r$ stars. A detailed discussion will be presented
in Bisterzo et al. (in preparation).

\section{Results}

\begin{figure}
\includegraphics[angle=270,width=18pc]{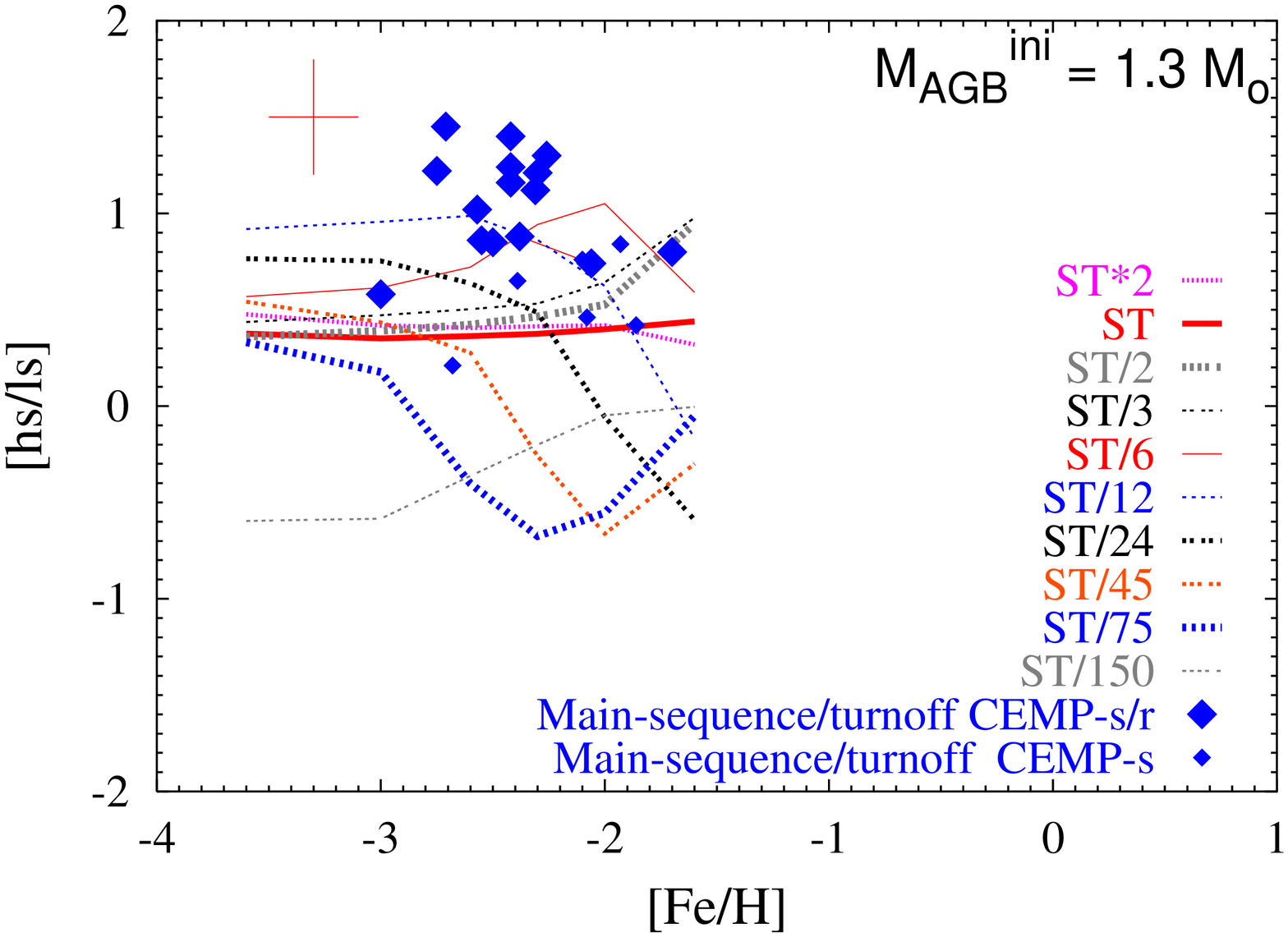}
\includegraphics[angle=270,width=18pc]{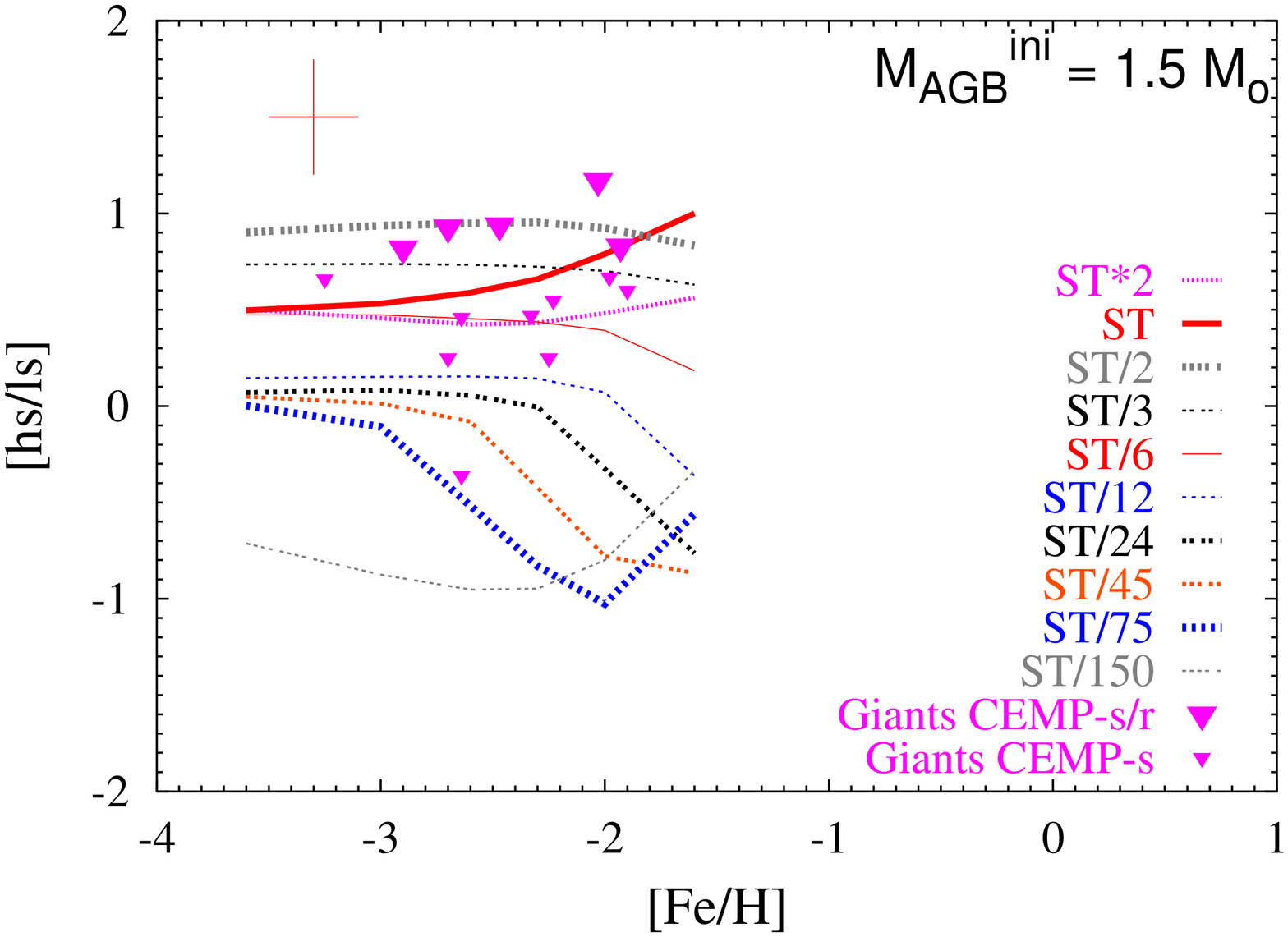}
\caption{Left panel: the [hs/ls] ratios observed in main-sequence/turnoff or subgiants CEMP-$s$
and CEMP-$s/r$ stars versus metallicity, compared with AGB models of initial 
mass $M$ = 1.3 $M_\odot$ and a range of $^{13}$C-pockets.
References are 
\cite{aoki02b,aoki06,aoki08,barklem05,behara10,cohen03,cohen06,ivans05,JB02,JB04,jonsell06,thompson08,tsangarides05}.
Right panel: the [hs/ls] ratios observed in CEMP-$s$ and CEMP-$s/r$ giants versus 
metallicity, compared with AGB models of initial mass $M$ = 1.5 $M_\odot$ and a 
range of $^{13}$C-pockets. Similar predictions are obtained by $M$ = 2 $M_\odot$ models. 
References are 
\cite{aoki02a,aoki06,barbuy05,barklem05,cohen06,GA10,goswami06,masseron06,roederer08,vaneck03}.
No initial $r$-process enhancement is assumed in both cases.}
\label{fig1}
\end{figure}

Among CEMP-$s$ stars in the literature, we selected only
those with Eu detection. About half of them are CEMP-$s/r$.
\\
In Fig.~\ref{fig1}, we analyse the $s$-process indicator [hs/ls] (where ls 
= Y, Zr and hs = La, Nd, Sm) versus metallicity, by comparing CEMP-$s/r$ 
and CEMP-$s$ observations with AGB models of initial mass $M$ = 1.3 
$M_\odot$ (left panel) or 1.5 $M_\odot$ (right panel). 
AGB models are described by \cite{gallino98} (updated by \cite{bisterzo10}). 
Starting from the case `ST' defined by \cite{gallino98,arlandini99}, a range 
of $^{13}$C-pockets is adopted by multiplying or dividing the $^{13}$C 
(and $^{14}$N) in the pocket by different factors. Theoretical lines in the 
Figure represent pure $s$-process AGB predictions for cases from `ST$\times$2' 
down to `ST/150'. 
For simplicity, in this preliminary analysis we distinguish between 
main-sequence/turnoff stars or subgiants having not suffered the first dredge-up 
(FDU) episode (diamonds in left panel), and late subgiants or
giants (down-rotated
triangles in right panel). The FDU involves about 80\% of the mass 
of the star \cite{busso99}. 
In case of binary systems with mass transfer like
CEMP-$s$ stars, this mixing strongly dilutes the C and $s$-rich material previously 
transferred from the AGB companion. This means that the [El/Fe] observed in a 
CEMP-$s$ giant is about 1 dex lower than in the envelope 
of the AGB companion\footnote{To simulate mixing processes we define the 
logarithmic ratio `dil' as
$\rm{dil = \log\left(\frac{M^{env}_{\star}}{\Delta M^{trans}_{AGB}}\right)}$,
where $M^{\rm env}_{\star}$ represents the mass of the convective 
envelope of the observed star before the mixing, 
and $\Delta M^{\rm trans}_{\rm AGB}$ is the AGB total mass transferred (see 
\cite{bisterzo10}).}.
CEMP-$s/r$ stars are represented by big symbols while CEMP-$s$ by little 
symbols. References are given in the caption of the Figures.
In average, CEMP-$s/r$ stars show higher [hs/ls] than CEMP-$s$. 
Moreover, some CEMP-$s/r$ have an observed [hs/ls] ratio higher than the
AGB predictions, but still compatible within the errorbars. 

\begin{figure}
\includegraphics[angle=270,width=18pc]{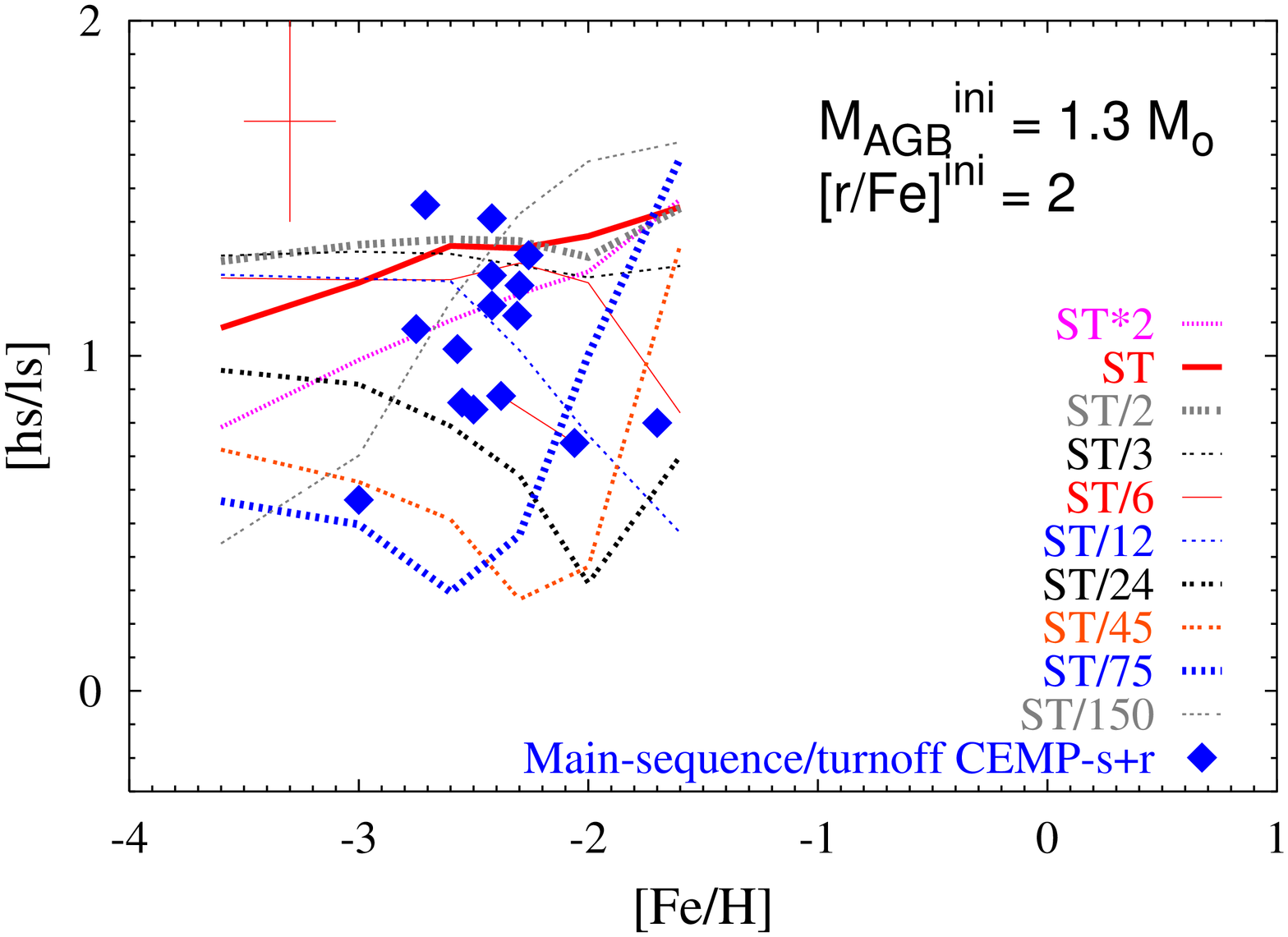}
\includegraphics[angle=270,width=18pc]{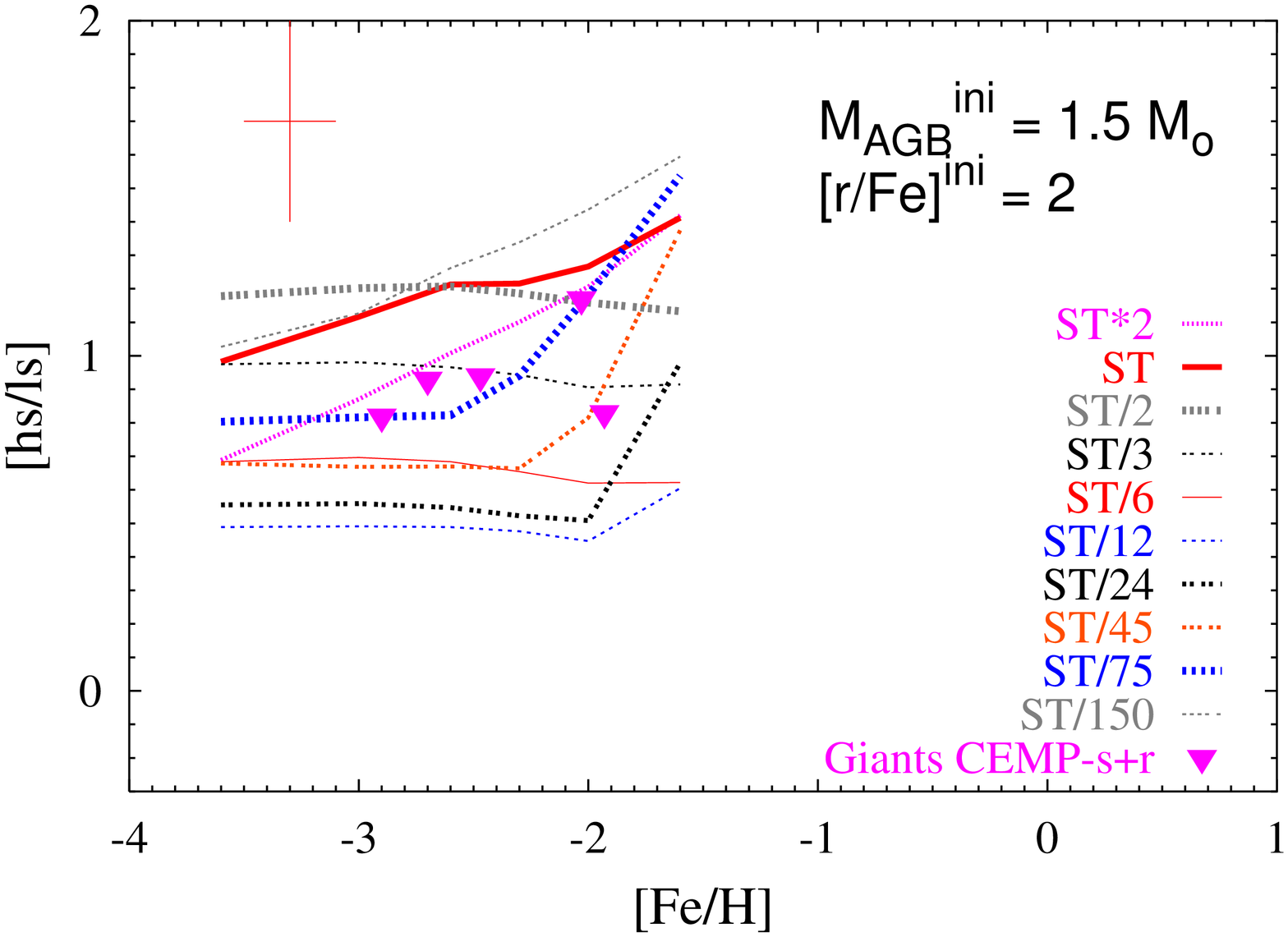}
\caption{Left panel: the [hs/ls] ratios observed in main-sequence/turnoff or subgiant CEMP-$s/r$
stars versus metallicity, compared with AGB models of initial mass $M$ = 1.3 $M_\odot$
and a range of $^{13}$C-pockets. 
Right panel: the [hs/ls] ratios observed in CEMP-$s/r$ giants versus metallicity, 
compared with AGB models of initial mass $M$ = 1.5 $M_\odot$ and a range of 
$^{13}$C-pockets. 
An initial $r$-enhancement of [r/Fe]$^{\rm ini}$ = 2.0 dex is adopted in both cases.
A dilution of 1 dex is applied to $M$ = 1.5 $M_\odot$ models, which best accounts 
for giant CEMP-$s$ or CEMP-$s/r$.}
\label{fig3}
\end{figure}

In Fig.~\ref{fig3}, the observed [hs/ls] ratio is compared with AGB models of 
initial masses $M$ = 1.3 $M_\odot$ (left panel) or 1.5 $M_\odot$ (right 
panel) with a high initial $r$-process enhancement, [r/Fe]$^{\rm ini}$ = 
2.0 dex (corresponding to [Eu/Fe] $^{\rm ini}$ = 2.0 dex). Only CEMP-$s/r$
 stars are shown in this Figure (big symbols).
The choice of the initial $r$-process contribution to heavy elements was 
based on the $r$-process solar predictions 
\cite{arlandini99,bisterzo10}. 
In particular, we applied an initial $r$-process contribution of 30\% to 
solar La, 40\% to solar Nd, and 70\% to solar Sm. 
In first approximation we assumed a solar-scaled 
Y and Zr.
This because the [Y,Zr/Fe] ratios observed in unevolved
halo stars reach maximum values of about 0.5 dex, which little affects
the [ls/Fe] in CEMP-$s$. 
The resulting maximum [hs/ls]$_{\rm s+r}$ with [r/Fe]$^{\rm ini}$ = 2.0 shown in
Fig.~\ref{fig3} is about 0.3 dex higher than the predicted [hs/ls]$_{\rm s}$
with no initial $r$-enhancement.
Note that the $s$-process index [hs/ls]$_{\rm s}$ is independent of 
the dilution factor if no initial $r$-enhancement is adopted (Fig.~\ref{fig1},
right panel). 
Instead, in case of a high initial $r$-enhancement, the dilution factor 
affects [hs/ls]$_{\rm s+r}$, because both stars belonging 
to the binary system are initially $r$-enriched. In Fig.~\ref{fig3}, 
right panel, a dil = 1.0 dex is applied.


%

%

\section{Conclusions}

To explain the origin of CEMP-$s/r$, we hypothesised that the molecular cloud 
from which the binary system formed was already enriched in $r$-process elements.
Subsequently, the $s$-process elements synthesised by the AGB companion 
are transferred by stellar winds on to the observed star.
The $s$-process nucleosynthesis is not affected by the initial $r$-enhancement
of the molecular cloud. 
However, for high $r$-process enrichment ([r/Fe]$^{\rm ini}$ = 2), one 
should account for the $r$-process contribution to solar La, Nd and Sm (30\%, 40\%, 70\%). 
In agreement with the [Y,Zr/Fe] observed in unevolved halo stars, we adopt solar 
scaled initial Y and Zr values.
This increases [hs/ls] by $\sim$ 0.3 dex. This is sustained by observations
in CEMP-$s/r$ stars, which show an [hs/ls] ratio in average higher than that 
observed in CEMP-$s$. Note that the [hs/ls] observed in CEMP-$s/r$ stars may
be in agreement with pure $s$-process predictions within the errorbars.
 A deeper analysis will be given in Bisterzo et al., in preparation.

\end{document}